\documentclass[12pt]{article}

\renewcommand{\thefootnote}{\fnsymbol{footnote}}
\usepackage{amsbsy,amssymb,latexsym,amsfonts,amsmath}
\usepackage{mathrsfs}
\usepackage{graphicx}
\usepackage{cite}
\usepackage{color}
\usepackage{bm}
\usepackage{here}
\numberwithin{equation}{section}
\newcommand{\bel}[1]{\begin{equation}\label{#1}}                     
\newcommand{\bal}[1]{\begin{eqnarray}\label{#1}}                     
\newcommand{\be}{\begin{equation}}
\newcommand{\ee}{\end{equation}}

\newcommand{\de}{\mathrm{d}}
\newcommand{\Tr}{\mathrm{Tr}}
\newcommand{\tr}{\mathrm{tr}}
\newcommand{\p}{\partial}

\newcommand{\mat}[1]{\begin{pmatrix} #1 \end{pmatrix}}

\newcommand{\bea}{\begin{equation}}
\newcommand{\eea}{\end{equation}}

\definecolor{red}{rgb}{1,0,0}
\definecolor{orange}{rgb}{1,0.5,0}
\definecolor{violet}{rgb}{0.7,0,1}

\def\cre{\color{red}}

\def\cg{\color{green}}
\def\cb{\color{blue}}

\def\b1{(-24,18)}\def\b2{(9,29)}\def\b3{(30,0)}\def\b4{(9,-29)}\def\b5{(-24,-18)}

\renewcommand{\b}{\beta}

\begin{document}
%%%%%%%%%%%%%%%%%%%%%%%%%%%%%%%%%%%%%%%%%%%%%%%%%%%%%%%%%%%%%%%%%%%%%%%%%%%%%%%%
%%%%%%%%%%%%%%%%%%%%%%%%%%%%%%%%%%%%%%%%%%%%%%%%%%%%%%%%%%%%%%%%%%%%%%%%%%%%%%%%%%%%%%%%%%
%
% title page
%
%%%%%%%%%%%%%%%%%%%%%%%%%%%%%%%%%%%%%%%%%%%%%%%%%%%%%%%%%%%%%%%%%%%%%%%%%%%%%%%%%%%%%%%%%
\begin{titlepage}
%%%%%%%%%%%%%%%%%%%% preprint # %%%%%%%%%%%%%%%%%%
\begin{flushright}
\normalsize
%\filename
~~~~
OCU-PHYS 498\\
NITEP 9\\
March, 2019\\
\end{flushright}
%%%%%%%%%%%%%%%%%%%%%%%%%%%%%%%%%%%%%%%%%%%%%%%%%%

\vspace{15pt}

%%%%%%%%%%%%%%%%%%%% title %%%%%%%%%%%%%%%%%%%%%%%
\begin{center}
{\LARGE Generalized cut operation } \\ 
{\LARGE associated with higher order variation in tensor models}
\end{center}
%%%%%%%%%%%%%%%%%%%%%%%%%%%%%%%%%%%%%%%%%%%%%%%%%%

\vspace{23pt}

%%%%%%%%%%%%%%%%%%% authors %%%%%%%%%%%%%%%%%%%%%%
\begin{center}
{ H. Itoyama$^{a, b,c}$\footnote{e-mail: itoyama@sci.osaka-cu.ac.jp}
  and  R. Yoshioka$^c$\footnote{e-mail: yoshioka@sci.osaka-cu.ac.jp}  }\\
%%%%%%%%%%%%%%%%%%%%%%%%%%%%%%%%%%%%%%%%%%%%%%%%%%
%
\vspace{18pt}
%
%%%%%%%%%%%%%%%%%%% affiliation %%%%%%%%%%%%%%%%%%%
$^a$\it Nambu Yoichiro Institute of Theoretical and Experimental Physics (NITEP),\\
Osaka City University\\
\vspace{5pt}

$^b$\it Department of Mathematics and Physics, Graduate School of Science,\\
Osaka City University\\
\vspace{5pt}

$^c$\it Osaka City University Advanced Mathematical Institute (OCAMI)

\vspace{5pt}

3-3-138, Sugimoto, Sumiyoshi-ku, Osaka, 558-8585, Japan \\

\end{center}
%%%%%%%%%%%%%%%%%%%%%%%%%%%%%%%%%%%%%%%%%%%%%%%%%%%
%
\vspace{20pt}
\begin{center}
Abstract\\
\end{center}
%%%%%%%%%%%%%%%%%%%% abstract %%%%%%%%%%%%%%%%%%%%% 
The cut and join operations play important roles in tensor models in general. We introduce a generalization of the cut operation associated with the higher order variations and demonstrate how they generate operators in the Aristotelian tensor model. We point out
that, by successive choices of appropriate variations, the cut operation generalized
this way can generate those operators which do not appear in the ring of the join operation, providing a tool to enumerate the operators by a level by level analysis recursively. We present  a set of rules that control the emergence of such operators.

%%%%%%%%%%%%%%%%%%%%%%%%%%%%%%%%%%%%%%%%%%%%%%%%%%%

\vfill

\end{titlepage}

%%%%%%%%%%%%%%%%%%%%
\renewcommand{\thefootnote}{\arabic{footnote}}
\setcounter{footnote}{0}
%%%%%%%%%%%%%%%%%%%%

%%%%%%%%%%%%%%%%%%%%%%%%%%%%%%%%%%%%%%%%%%%%%%%%%%%%%%%%%%%%%%%%%%%%%%%%%%%%%%%%%%
%%%%%%%%%%%%%%%%%%%%%%%%%%%%%%%%%%%%%%%%%%%%%%%%%%%%%%%%%%%%%%%%%%%%%%%%%%%%%%%%%%
\section{Introduction}
%%%%%%%%%%%%%%%%%%%%%%%%%%%%%%%%%%%%%%%%%%%%%%%%%%%%%%%%%%%%%%%%%%%%%%%%%%%%%%%%%%
%%%%%%%%%%%%%%%%%%%%%%%%%%%%%%%%%%%%%%%%%%%%%%%%%%%%%%%%%%%%%%%%%%%%%%%%%%%%%%%%%%
The tensor model has a long history of its research \cite{David1985,KMK1985,ADJ1991,Sasakura1991,Ginsparg1991,Gross1992}.  
Recent interest has come from 
an intersection of thoughts on holography and randomness which are realized by several
phenomena in quantum gravity in lower dimensions  \cite{Witten1610,Gurau1611,Gurau1702,CT1512,KT1611,GR1610,GR1702,GR1706,GR1710,KSS1612,BC1304,Ferrari1701,BLT1702,Gurau1203,Bonzom1208,IMM1703,IMM1704,IMM1710,IMM1808,MM1706}.
 Some of the insights come from the Virasoro structure of the matrix models \cite{David1990,MM1990,AM1990,IM1991N,IM1991W,DVV1991,FKN1991,FKN1992} and its combinatorics and finding its counterpart in tensor models in general is expected to pave the way to bring progress in this field. Recent references include 
 \cite{MM1705,MM1807,Morozov1810,Morozov1901,BGRR1105,Gurau1105,BGR1307,DR1706,dMKGT1707,BGR1708,Alexandrov1608,Alexandrov1412,Alexandrov1009,WWYZZ1901}.

The Virasoro algebra has a natural extension named $w_{1+\infty}$ algebra whose
 roles at the 2-dimensional gravity and some integrable models have been well investigated.
In particular, the constraints of $w_{1+\infty}$ type coming from the higher order contribution of the variation \cite{IM1991W} have turned out to be algebraically independent and nontrivial in some of the matrix models such as the two-matrix model. Here we would like to discuss such higher order contributions arising from the change of the integration measure under the variation.

The basic structure of the contributions from the action 
 is the join operation defined by       
\begin{equation} \label{join} 
 \{K, K' \} 
 = \frac{\partial K}{\partial A_{a_1a_2\cdots a_r}} 
 \frac{\partial K'}{\partial \bar{A}^{a_1a_2\cdots a_r}},  
\end{equation}
where $K$ and $K'$ are arbitrary operators and 
 the summation over the repeated indices is implied.  
When we choose appropriate keystone operators for $K$ and $K'$, 
 a block of independent operators called join pyramid 
 is successively generated by the join operation. 
In other words, the join operation forms a ring 
 whose elements are independent operators and
 the multiplication is given by \eqref{join}.  
There are, however, operators which are not involved in the join pyramid and 
 these can not be ignored 
 because the cut operation which underlies the contribution 
 from the variation of the integration measure generates these. 
 These pieces of structure were discovered in \cite{IMM1710}. 
 (In contrast, the cut and join operation in one matrix model, namely, $r=2$, is depicted as going up and down at integer points on one-dimensional half-line.)
The cut operation is defined by 
\begin{equation} \label{cut}
 \Delta K = \frac{\partial^2 K}{\partial A_{a_1a_2\cdots a_r} 
 \partial \bar{A}^{a_1a_2\cdots a_r}}, 
\end{equation}
and corresponds to going up one stair (by one level) in the join pyramid.   
There is no systematic way to predict when a new operator appears and,  
 in the situation of \cite{IMM1710},  
one can try to discover this by acting the cut operation on all operators in the join pyramid only. 

Taking the above mentioned role of the cut and join operation into account, 
we expect that the cut operation plays an important role in resolving
  the enumeration problem of the operators in tensor models.  
Below we will investigate higher order contributions 
 to the constraints from the variation of the integral measure. 

This paper is organized as follows: 
In section 2, the higher order variations of the integration measure are considered. 
In section 3, we discuss the successive choices of the variations.
In section 4, we check that our choice of the variation is correct  
 up to the level 6 operators.
In section 5, a procedure of generating   the operators not included in the join pyramid is
 described.    
 
We owe most of our terminology to those seen in \cite{IMM1703,IMM1704,IMM1710}.

%%%%%%%%%%%%%%%%%%%%%%%%%%%%%%%%%%%%%%%%%%%%%%%%%%%%%%%%%%%%%%%%%%%%%%%%%%%%%%%%%%
%%%%%%%%%%%%%%%%%%%%%%%%%%%%%%%%%%%%%%%%%%%%%%%%%%%%%%%%%%%%%%%%%%%%%%%%%%%%%%%%%%
\section{Higher order variation}
%%%%%%%%%%%%%%%%%%%%%%%%%%%%%%%%%%%%%%%%%%%%%%%%%%%%%%%%%%%%%%%%%%%%%%%%%%%%%%%%%%
%%%%%%%%%%%%%%%%%%%%%%%%%%%%%%%%%%%%%%%%%%%%%%%%%%%%%%%%%%%%%%%%%%%%%%%%%%%%%%%%%%
Let us consider the rank $r=3$ Aristotelian tensor model. 
Let $A$ be a rank 3 tensor with its component $A_{\cre{a_1}\cg{a_2}\cb{a_3}}$ 
 and be its conjugate $\bar{A}$ with $\bar{A}^{\cre{a_1}\cg{a_2}\cb{a_3}}$. 
Each index $a_i$, $i=1,2,3$ runs over $1 , \cdots, N_i$ 
and is colored respectively in red, green and blue.  
The shift of integration variables of the partition function
  is defined by  $A \to A + \delta A$ and $\bar{A} \to \bar{A} + \delta \bar{A}$ with  
\begin{equation}\label{variation}
 \delta A_{\cre{a_1}\cg{a_2}\cb{a_3}} = \frac{\p K}{\p \bar{A}^{\cre{a_1}\cg{a_2}\cb{a_3}}}, ~~~~~~~~~
  \delta \bar{A}^{\cre{a_1}\cg{a_2}\cb{a_3}} 
  = \frac{\p K}{\p {A}_{\cre{a_1}\cg{a_2}\cb{a_3}}}, 
\end{equation}
for arbitrary $K$. 
%Then
%\begin{equation}
% \de \mat{\delta A \\ \delta \bar{A}} = F \mat{\de A \\ \de \bar{A}},
%\end{equation}
As the line element is given by  
\begin{equation}
 \de s^2_{A} = \sum_{\cre{a_1}, \cg{a_2}, \cb{a_3}} 
 \de A_{\cre{a_1}\cg{a_2}\cb{a_3}} \de \bar{A}^{\cre{a_1}\cg{a_2}\cb{a_3}}
 = \frac{1}{2} \mat{\de \bar{A} & \de A} \mat{\de A \\ \de \bar{A}}, 
\end{equation}
its response under \eqref{variation} is 
\begin{equation}
 \de s^2_{A+\delta A} = 
 \mat{\de \bar{A} & \de A} ( 1 + F )^2 \mat{\de A \\ \de \bar{A}}, 
\end{equation}
where  the matrix $F$ of size $(2 N_1N_2N_3) \times (2 N_1N_2N_3)$  is defined by
\begin{equation}
F = \mat{\p\bar{\p} K & \bar{\p}\bar{\p} K \\ \p\p K & {\p}\bar{\p} K}, ~~~~~~~~
\p = \frac{\p}{\p A},~~~~~ \bar{\p} = \frac{\p}{\p \bar{A}}.
\end{equation}
The measure is, therefore, transformed as 
\begin{equation}
 [\de A] [\de \bar{A}] ~~ \longrightarrow ~~
 \det(1+F) [\de A] [\de \bar{A}],
\end{equation}
where 
\begin{align}
 \det(1+F) =  1 + \sum_{n=1}^{\infty} \frac{(-1)^n}{n} \tr F^n 
 + \frac{1}{2} \sum_{n,m} \frac{(-1)^{n+m}}{nm} \tr F^n \tr F^m + \cdots. 
\end{align}

The cut operator $\Delta$ \eqref{cut} corresponds to    
\begin{equation}
 \tr F = 2 \Delta K.  
\end{equation}
We are interested in the higher order contribution at the response of the measure
 under the general variation 
and the gauge-invariant operators which are contained in $\det (1+F)$.
We use the pictorial representation of the operators as follows:
The tensor $A$ (resp. $\bar{A}$) are denoted by a white circle (resp. a black dot)
and the contractions of indices are denoted by colored lines connecting between 
 the white circles and the black dots. 
For example, 
\be 
 K_1 \equiv A_{{\cre a_1} {\cg a_2} {\cb a_3}} \bar{A}^{{\cre a_1} {\cg a_2} {\cb a_3}}
 =  
\setlength{\unitlength}{0.13cm}
\begin{picture}(14,7)
  \linethickness{0.2mm}
  \put(2,1){\cb \line(1,0){10}}
  \put(-8,5){{\cre \qbezier(10,-4)(15,0)(20,-4)}} 
  \put(-8,5){{\cg \qbezier(10,-4)(15,-8)(20,-4)}}
  \put(2.,1){\circle{2}}
  \put(12.,1){\circle*{2}}
  \put(0,-4){\mbox{$A$}}
  \put(11,-4){\mbox{$\bar{A}$}}
\end{picture}
\ee\\
The connected operators come from $\tr F^n$. 
The number of $A$ in the operator under consideration is called level of the operator. 
In the case of $n=2$ and higher, 
the generalized cut operation $\tr F^n$ raises the level and 
therefore it can be used as the procedure which generates the higher level operators, 
while the usual cut operation \eqref{cut} lowers the level of the operators by one.   
In the next section, 
 we see that all connected operators at each level are included in $\tr F^n$
 if $K$ is appropriately chosen. 

%%%%%%%%%%%%%%%%%%%%%%%%%%%%%%%%%%%%%%%%%%%%%%%%%%%%%%%%%%%%%%%%%%%%%%%%%%%%%%%%%%
%%%%%%%%%%%%%%%%%%%%%%%%%%%%%%%%%%%%%%%%%%%%%%%%%%%%%%%%%%%%%%%%%%%%%%%%%%%%%%%%%%
\section{Choice of $K$}
%%%%%%%%%%%%%%%%%%%%%%%%%%%%%%%%%%%%%%%%%%%%%%%%%%%%%%%%%%%%%%%%%%%%%%%%%%%%%%%%%%
%%%%%%%%%%%%%%%%%%%%%%%%%%%%%%%%%%%%%%%%%%%%%%%%%%%%%%%%%%%%%%%%%%%%%%%%%%%%%%%%%%
In this section, we seek for the appropriate choice of the variation $K$  
 to construct all operators. 
Now let us choose temporarily 
\begin{equation}\label{K<2}
 K_{\leq 2} = K_{\cre{2}} + K_{\cg{2}} + K_{\cb{2}},
\end{equation} 
where 
\be
K_{\cre{2}} = A_{\cre{a_1}\cg{a_2}\cb{a_3}} A_{\cre{a'_1}\cg{a'_2}\cb{a'_3}} 
\bar{A}^{\cre{a_1}\cg{a'_2}\cb{a'_3}}\bar{A}^{\cre{a'_1}\cg{a_2}\cb{a_3}} = 
\setlength{\unitlength}{0.1cm}
\begin{picture}(20,10)
  \linethickness{0.2mm}
  \put(0,10){{\cg \qbezier(6,-4)(13,0)(20,-4)}} 
  \put(0,10){{\cb \qbezier(6,-4)(13,-8)(20,-4)}}
  \put(6,-4){\cre \line(0,1){10}}
  \put(20,-4){\cre \line(0,1){10}}
  \put(6.,6){\circle{2}}
  \put(20.,6){\circle*{2}}
  \put(0,0){{\cg \qbezier(6,-4)(13,0)(20,-4)}} 
  \put(0,0){{\cb \qbezier(6,-4)(13,-8)(20,-4)}}
  \put(6.,-4){\circle*{2}}
  \put(20.,-4){\circle{2}} 
  %\put(8,-9){\mbox{$A$}}
  %\put(18,-9){\mbox{$\bar{A}$}}
\end{picture}~~. 
\ee \\   
The operator \eqref{K<2} is the linear combination of the level 2 operators 
 and all operators in $\tr F^n$ are level $k=n$.   
Although $\tr F^n$ consists of $\p\bar{\p}K_{\leq 2}$ , $\p\p K_{\leq 2}$ 
 and $\bar{\p}\bar{\p} K_{\leq 2}$, 
it turns out that only $\p\bar{\p} K_{\leq 2}$ is necessary below.   
Pictorially, 
\begin{align}
 &\p\bar{\p} K_{{\cre 2}} = 2 ( A_{\cre{a} \cg{a_2} \cb{a_3}} 
 \bar{A}^{\cre{a} \cg{\bar{a}_2} \cb{\bar{a}_3}} \delta_{\cre{a_1}}^{\cre{\bar{a}_1}}
 + A_{\cre{a_1} \cg{a} \cb{b}} 
 \bar{A}^{\cre{\bar{a}_1} \cg{a} \cb{b}} \delta_{\cg{a_2}}^{\cg{\bar{a}_2}} 
 \delta_{\cb{a_3}}^{\cb{\bar{a}_3}})
 \equiv 2 \biggl\{ ({\cre 2}) + ({\cre 2})' \biggl\}, \\
 &\p\bar{\p} K_{{\cg 2}} = 2 ( A_{\cre{a_1} \cg{a} \cb{a_3}} 
 \bar{A}^{\cre{\bar{a}_1} \cg{{a}} \cb{\bar{a}_3}} \delta_{\cg{a_2}}^{\cg{\bar{a}_2}}
 + A_{\cre{b} \cg{a_2} \cb{a}} 
 \bar{A}^{\cre{{b}} \cg{\bar{a}_2} \cb{a}} \delta_{\cre{a_1}}^{\cre{\bar{a}_1}} 
 \delta_{\cb{a_3}}^{\cb{\bar{a}_3}})
 \equiv 2 \biggl\{ ({\cg 2}) + ({\cg 2})' \biggl\}, \\
 &\p\bar{\p} K_{{\cb 2}} = 2 ( A_{\cre{a_1} \cg{a_2} \cb{a}} 
 \bar{A}^{\cre{\bar{a}_1} \cg{\bar{a}_2} \cb{{a}}} \delta_{\cb{a_3}}^{\cb{\bar{a}_3}}
 + A_{\cre{a} \cg{b} \cb{a_3}} 
 \bar{A}^{\cre{{a}} \cg{b} \cb{\bar{a_3}}} \delta_{\cre{a_1}}^{\cre{\bar{a}_1}} 
 \delta_{\cg{a_2}}^{\cg{\bar{a}_2}})
 \equiv 2 \biggl\{ ({\cb 2}) + ({\cb 2})' \biggl\}, 
\end{align}
where 
\begin{align}
&({\cre 2}) = ~ 
\setlength{\unitlength}{0.1cm}
\begin{picture}(40,10)
  \linethickness{0.2mm}
  \put(5,6){\cre \line(1,0){30}}
  \put(5,-4){\cg \line(1,0){10}}
  \put(15,-4){\cre \line(1,0){10}}
  \put(25,-4){\cg \line(1,0){10}} 
  \put(5,0){{\cb \qbezier(0,1)(10,0)(10,-4)}} 
  \put(5,0){{\cb \qbezier(30,1)(20,0)(20,-4)}} 
  \put(15.,-4){\circle{2}}
  \put(25.,-4){\circle*{2}}
  \put(13,-9){\mbox{$A$}}
  \put(23,-9){\mbox{$\bar{A}$}}
  \put(0,5){\mbox{$\cre{a_1}$}}
  \put(0,0){\mbox{$\cb{a_3}$}}
  \put(0,-5){\mbox{$\cg{a_2}$}}
  \put(36,5){\mbox{$\cre{a'_1}$}}
  \put(36,0){\mbox{$\cb{a'_3}$}}
  \put(36,-5){\mbox{$\cg{a'_2}$}}
  \put(19,-3){\mbox{$\cre a$}}
\end{picture} 
&({\cre 2})' = ~ 
\setlength{\unitlength}{0.1cm}
\begin{picture}(40,10)
  \linethickness{0.2mm}
  \put(5,6){\cg \line(1,0){30}}
  \put(5,1){\cb \line(1,0){30}}
  \put(5,-4){\cre \line(1,0){10}}
  \put(25,-4){\cre \line(1,0){10}}
  \put(5,0){{\cg \qbezier(10,-4)(15,0)(20,-4)}} 
  \put(5,0){{\cb \qbezier(10,-4)(15,-8)(20,-4)}}
  \put(15.,-4){\circle{2}}
  \put(25.,-4){\circle*{2}}
  \put(13,-9){\mbox{$A$}}
  \put(23,-9){\mbox{$\bar{A}$}}
  \put(0,5){\mbox{$\cg{a_2}$}}
  \put(0,0){\mbox{$\cb{a_3}$}}
  \put(0,-5){\mbox{$\cre{a_1}$}}
  \put(36,5){\mbox{$\cg{a'_2}$}}
  \put(36,0){\mbox{$\cb{a'_3}$}}
  \put(36,-5){\mbox{$\cre{a'_1}$}}
  \put(19,-4.5){\mbox{$\cg a$}}
  \put(19,-9.5){\mbox{$\cb b$}}
\end{picture} \\
&({\cg 2}) = ~ 
\setlength{\unitlength}{0.1cm}
\begin{picture}(30,20)
  \linethickness{0.2mm}
  \put(5,6){\cg \line(1,0){30}}
  \put(5,-4){\cb \line(1,0){10}}
  \put(15,-4){\cg \line(1,0){10}}
  \put(25,-4){\cb \line(1,0){10}} 
  \put(5,0){{\cre \qbezier(0,1)(10,0)(10,-4)}} 
  \put(5,0){{\cre \qbezier(30,1)(20,0)(20,-4)}} 
  \put(15.,-4){\circle{2}}
  \put(25.,-4){\circle*{2}}
  \put(13,-9){\mbox{$A$}}
  \put(23,-9){\mbox{$\bar{A}$}}
  \put(0,5){\mbox{$\cg{a_2}$}}
  \put(0,0){\mbox{$\cre{a_1}$}}
  \put(0,-5){\mbox{$\cb{a_3}$}}
  \put(36,5){\mbox{$\cg{a'_2}$}}
  \put(36,0){\mbox{$\cre{a'_1}$}}
  \put(36,-5){\mbox{$\cb{a'_3}$}}
  \put(19,-3){\mbox{$\cg a$}}
\end{picture} 
&({\cg 2})' = ~ 
\setlength{\unitlength}{0.1cm}
\begin{picture}(40,20)
  \linethickness{0.2mm}
  \put(5,6){\cb \line(1,0){30}}
  \put(5,1){\cre \line(1,0){30}}
  \put(5,-4){\cg \line(1,0){10}}
  \put(25,-4){\cg \line(1,0){10}}
  \put(5,0){{\cb \qbezier(10,-4)(15,0)(20,-4)}} 
  \put(5,0){{\cre \qbezier(10,-4)(15,-8)(20,-4)}}
  \put(15.,-4){\circle{2}}
  \put(25.,-4){\circle*{2}}
  \put(13,-9){\mbox{$A$}}
  \put(23,-9){\mbox{$\bar{A}$}}
  \put(0,5){\mbox{$\cb{a_3}$}}
  \put(0,0){\mbox{$\cre{a_1}$}}
  \put(0,-5){\mbox{$\cg{a_2}$}}
  \put(36,5){\mbox{$\cb{a'_3}$}}
  \put(36,0){\mbox{$\cre{a'_1}$}}
  \put(36,-5){\mbox{$\cg{a'_2}$}}
  \put(19,-4.5){\mbox{$\cb a$}}
  \put(19,-9.5){\mbox{$\cre b$}}
\end{picture}\\
&({\cb 2}) = ~ 
\setlength{\unitlength}{0.1cm}
\begin{picture}(40,20)
  \linethickness{0.2mm}
  \put(5,6){\cb \line(1,0){30}}
  \put(5,-4){\cre \line(1,0){10}}
  \put(15,-4){\cb \line(1,0){10}}
  \put(25,-4){\cre \line(1,0){10}} 
  \put(5,0){{\cg \qbezier(0,1)(10,0)(10,-4)}} 
  \put(5,0){{\cg \qbezier(30,1)(20,0)(20,-4)}} 
  \put(15.,-4){\circle{2}}
  \put(25.,-4){\circle*{2}}
  \put(13,-9){\mbox{$A$}}
  \put(23,-9){\mbox{$\bar{A}$}}
  \put(0,5){\mbox{$\cb{a_3}$}}
  \put(0,0){\mbox{$\cg{a_2}$}}
  \put(0,-5){\mbox{$\cre{a_1}$}}
  \put(36,5){\mbox{$\cb{a'_3}$}}
  \put(36,0){\mbox{$\cg{a'_2}$}}
  \put(36,-5){\mbox{$\cre{a'_1}$}}
  \put(19,-3){\mbox{$\cb a$}}
\end{picture} 
&({\cb 2})' = ~ 
\setlength{\unitlength}{0.1cm}
\begin{picture}(40,20)
  \linethickness{0.2mm}
  \put(5,6){\cre \line(1,0){30}}
  \put(5,1){\cg \line(1,0){30}}
  \put(5,-4){\cb \line(1,0){10}}
  \put(25,-4){\cb \line(1,0){10}}
  \put(5,0){{\cre \qbezier(10,-4)(15,0)(20,-4)}} 
  \put(5,0){{\cg \qbezier(10,-4)(15,-8)(20,-4)}}
  \put(15.,-4){\circle{2}}
  \put(25.,-4){\circle*{2}}
  \put(13,-9){\mbox{$A$}}
  \put(23,-9){\mbox{$\bar{A}$}}
  \put(0,5){\mbox{$\cre{a_1}$}}
  \put(0,0){\mbox{$\cg{a_2}$}}
  \put(0,-5){\mbox{$\cb{a_3}$}}
  \put(36,5){\mbox{$\cre{a'_1}$}}
  \put(36,0){\mbox{$\cg{a'_2}$}}
  \put(36,-5){\mbox{$\cb{a'_3}$}}
  \put(19,-4.5){\mbox{$\cre a$}}
  \put(19,-9.5){\mbox{$\cg b$}}
\end{picture} 
\end{align}\\\\ 
In the subsections in what follows, we will show that all operators 
 at the first few levels denoted generically by $k$ are included in $\tr F^n$.

%%%%%%%%%%%%%%%%%%%%%%%%%%%%%%%%%%%%%%%%%%%%%%%%%%%%%%%%%%%%%%%%%%%%%%%%%%%%%%%%%%
\subsection{level $k=1$}
%%%%%%%%%%%%%%%%%%%%%%%%%%%%%%%%%%%%%%%%%%%%%%%%%%%%%%%%%%%%%%%%%%%%%%%%%%%%%%%%%%
The only connected operator is 
$K_1 = A_{\cre{a_1}\cg{a_2}\cb{a_3}} \bar{A}^{\cre{a_1}\cg{a_2}\cb{a_3}}$.
In the case of $n=1$, 
 $\tr F$ is the cut operation itself as mentioned above and we have
\begin{equation}
 \tr F = 4(N_1+N_2+N_3 + (N_1N_2+N_2N_3+N_3N_1)) K_1.
\end{equation}
Conversely, we can obtain $K_1$ as the form of, for example,   
\begin{equation}
 K_1  
 = \frac{1}{N_1} \Tr ({\cre 2}).
\end{equation}
Here the trace ``$\Tr$" denotes the contraction of all indices.
In the pictorial representation, it corresponds to connecting the two open lines 
 with the same color on the both sides.

%%%%%%%%%%%%%%%%%%%%%%%%%%%%%%%%%%%%%%%%%%%%%%%%%%%%%%%%%%%%%%%%%%%%%%%%%%%%%%%%%%
\subsection{level $k=2$}
%%%%%%%%%%%%%%%%%%%%%%%%%%%%%%%%%%%%%%%%%%%%%%%%%%%%%%%%%%%%%%%%%%%%%%%%%%%%%%%%%%
All connected operators are listed in the appendix A2 of \cite{IMM1710}. 
Similarly to the case of level $k=1$, $\tr F^2$ contains
\begin{align}
 &K_{\cre{2}} = \frac{1}{N_1} \Tr ({\cre 2})^2. 
\end{align}
The operators $K_{\cg{2}}$ and $K_{\cb{2}}$ are also obtained in a similar way.  

%%%%%%%%%%%%%%%%%%%%%%%%%%%%%%%%%%%%%%%%%%%%%%%%%%%%%%%%%%%%%%%%%%%%%%%%%%%%%%%%%%
\subsection{level $k=3$}
%%%%%%%%%%%%%%%%%%%%%%%%%%%%%%%%%%%%%%%%%%%%%%%%%%%%%%%%%%%%%%%%%%%%%%%%%%%%%%%%%%
All connected operators are listed in the appendix A3 of \cite{IMM1710}.
At $n=3$, $\tr F^3$ contains not only
\begin{align}
 &K_{\cre{3}} = \frac{1}{N_1} \Tr ({\cre 2})^3, \\
 &K_{\cre{2}\cg{2}} = \Tr ({\cre 2})^2 ({\cb 2}), 
\end{align}
but also 
\begin{align}
 K_{3W} &= \Tr ({\cre 2}) ({\cg 2}) ({\cb 2}) \cr
 &= \Tr ~
 \setlength{\unitlength}{0.1cm}
\begin{picture}(120,10)
  \linethickness{0.2mm}
  \put(0,6){\cre \line(1,0){20}}
  \put(0,-4){\cg \line(1,0){10}}
  \put(10,-4){\cre \line(1,0){10}}
  \put(20,-4){\cb \line(1,0){20}} 
  \put(0,0){{\cre \qbezier(20,6)(30,1)(40,-4)}}
  \put(0,0){{\cb \qbezier(0,1)(10,0)(10,-4)}} 
  \put(0,0){{\cg \qbezier(20,-4)(30,1)(40,6)}} 
  \put(10,-4){\circle{2}}
  \put(20,-4){\circle*{2}}  %
  \put(40,6){\cg \line(1,0){10}}
  \put(40,-4){\cg \line(1,0){10}}
  \put(50,-4){\cre \line(1,0){10}}  
  {\cb \qbezier(50,-4)(60,1)(70,6)} 
  \put(40,-4){\circle{2}}
  \put(50,-4){\circle*{2}}
  \put(70,6){\cb \line(1,0){20}}
  \put(60,-4){\cre \line(1,0){10}}
  \put(70,-4){\cb \line(1,0){10}}
  \put(80,-4){\cre \line(1,0){10}} 
  {\cg \qbezier(50,6)(60,1)(70,-4)} 
  \put(60,0){{\cg \qbezier(30,1)(20,0)(20,-4)}} 
  \put(70,-4){\circle{2}}
  \put(80,-4){\circle*{2}}
  \put(9,-9){1}
  \put(19,-9){$\bar{1}$}
  \put(39,-9){3}
  \put(49,-9){$\bar{2}$}
  \put(69,-9){2}
  \put(79,-9){$\bar{3}$}
\end{picture} \cr
&= ~~~~
 \setlength{\unitlength}{0.03cm}
\begin{picture}(100,80)
  \linethickness{0.2mm}
  \put(-4,0){
  {\cre{\qbezier(20,34)(10,17)(0,0)}
  {\qbezier(20,-34)(40,-34)(60,-34)}
  {\qbezier(80,0)(70,17)(60,34)}}}
  \put(-4,0){
  {\cg 
  {\qbezier(0,0)(10,-17)(20,-34)}
  {\qbezier(60,-34)(70,-17)(80,0)}
  {\qbezier(60,34)(40,34)(20,34)}}}
  \put(-4,0){
  {\cb 
  {\qbezier(20,34)(40,0)(60,-34)}
  {\qbezier(0,0)(40,0)(80,0)}
  {\qbezier(20,-34)(40,0)(60,34)}}}
  \put(20,34){\circle{6}}
  \put(20,-34){\circle{6}}
  \put(80,0){\circle{6}}
  \put(0,0){\circle*{6}}
  \put(60,-34){\circle*{6}}
  \put(60,34){\circle*{6}}
  \put(5,34){1}
  \put(5,-39){2}
  \put(90,-5){3}
  \put(-15,-5){$\bar{1}$}
  \put(70,-39){$\bar{2}$}
  \put(70,34){$\bar{3}$}
  \end{picture}. \label{3W}
\end{align}\\\\
The last one $K_{3W}$ cannot be obtained by the join operation. 
Hereafter, and in \cite{IMM1703,IMM1704,IMM1710} such operators are called  
 secondary operators.  
In the original procedure of \cite{IMM1703,IMM1704,IMM1710}, 
 we had to act the  original cut operation \eqref{cut} on all of the level $k=4$ operators 
 in order to discover the secondary operator $K_{3W}$. 

%%%%%%%%%%%%%%%%%%%%%%%%%%%%%%%%%%%%%%%%%%%%%%%%%%%%%%%%%%%%%%%%%%%%%%%%%%%%%%%%%%
\subsection{level $k=4$}
%%%%%%%%%%%%%%%%%%%%%%%%%%%%%%%%%%%%%%%%%%%%%%%%%%%%%%%%%%%%%%%%%%%%%%%%%%%%%%%%%%
The independent operators are listed in the appendix A4 of \cite{IMM1710}.
At $n=4$, $\tr F^4$ contains
\begin{align}
 &K_{\cre 4} = \frac{1}{N_1} \Tr ({\cre 2})^4, \\
 &K_{{\cb 3}{\cg 2}} =  \frac{1}{N_1} \Tr ({\cre 2})^3 ({\cg 2})', \\
 &K_{222} = \frac{1}{N_1} \Tr ({\cre 2})^2 ({\cg 2})' ({\cb 2})', \\
 &K_{{\cre 2}{\cg 2}{\cre 2}} 
 = \frac{1}{N_2} \Tr ({\cre 2})' ({\cg 2}) ({\cre 2})' ({\cg 2}), \\
 &K_{{\cre 2}{\cb 2}{\cg 2}} 
 = \frac{1}{N_3} \Tr ({\cre 2})' ({\cb 2}) ({\cg 2})' ({\cb 2}), \\
 \checkmark~~~~
 &K_{4C} = \Tr ({\cre 2}) ({\cg 2}) ({\cre 2}) ({\cg 2}), \label{4C}\\
 &K_{{\cre 3}{\cre 1}W} = \Tr ({\cre 2})^2 ({\cb 2}) ({\cg 2}), \\
 \checkmark~~~~ 
 &K_{{\cre 2}{\cre 2}W} 
 = \Tr ({\cre 2}) ({\cg 2}) ({\cre 2}) ({\cb 2}). \label{22W} 
\end{align}
The two operators $K_{4C}$ and $K_{\cre{22}W}$ denoted by $\checkmark$ 
 are secondary operators. 
We adopt $\checkmark$ notation to indicate secondary.

%%%%%%%%%%%%%%%%%%%%%%%%%%%%%%%%%%%%%%%%%%%%%%%%%%%%%%%%%%%%%%%%%%%%%%%%%%%%%%%%%%
\subsection{level $k=5$}
%%%%%%%%%%%%%%%%%%%%%%%%%%%%%%%%%%%%%%%%%%%%%%%%%%%%%%%%%%%%%%%%%%%%%%%%%%%%%%%%%%
At level $k=5$, 
$K_{\rm XXV}$, $K_{\rm XXVI}$ and $K_{\rm XXVIII}$ are still missing even
 with the generalized cut operation of this paper by the choice \eqref{K<2}.
In order to resolve this, let us replace \eqref{K<2} by 
\begin{equation}\label{K<3}
 K_{\leq 3} = K_{\cre{2}} + K_{\cg{2}} + K_{\cb{2}} + K_{3W}. 
\end{equation} 
In this case, we have, in addition,   
\begin{equation}
 \p\bar{\p} K_{3W} = 3\biggl\{ (3W)_{\cre r} + (3W)_{\cg g} + (3W)_{\cb b} \biggl\},
\end{equation}
where
\begin{align}
 &(3W)_{\cre r} = 
 \setlength{\unitlength}{0.1cm}
\begin{picture}(50,10)
  \linethickness{0.2mm}
  \put(0,6){\cre \line(1,0){50}}
  \put(0,-4){\cg \line(1,0){10}}
  \put(10,-4){\cre \line(1,0){10}}
  \put(20,-4){\cg \line(1,0){10}} 
  \put(30,-4){\cre \line(1,0){10}}
  \put(40,-4){\cg \line(1,0){10}}
  \put(0,0){{\cb \qbezier(0,1)(30,0)(30,-4)}} 
  \put(0,0){{\cb \qbezier(50,1)(20,0)(20,-4)}} 
  \put(0,0){{\cb \qbezier(10,-4)(25,-15)(40,-4)}}
  \put(10.,-4){\circle{2}}
  \put(20.,-4){\circle*{2}}
  \put(30.,-4){\circle{2}}
  \put(40.,-4){\circle*{2}}
\end{picture} ~~=~~
\begin{picture}(50,10)
  \linethickness{0.2mm}
  \put(0,6){\cre \line(1,0){50}}
  \put(0,-4){\cb \line(1,0){10}}
  \put(10,-4){\cre \line(1,0){10}}
  \put(20,-4){\cb \line(1,0){10}} 
  \put(30,-4){\cre \line(1,0){10}}
  \put(40,-4){\cb \line(1,0){10}}
  \put(0,0){{\cg \qbezier(0,1)(30,0)(30,-4)}} 
  \put(0,0){{\cg \qbezier(50,1)(20,0)(20,-4)}} 
  \put(0,0){{\cg \qbezier(10,-4)(25,-15)(40,-4)}}
  \put(10.,-4){\circle{2}}
  \put(20.,-4){\circle*{2}}
  \put(30.,-4){\circle{2}}
  \put(40.,-4){\circle*{2}}
\end{picture}~~, ~~ \\ 
 &(3W)_{\cg g} = 
 \setlength{\unitlength}{0.1cm}
\begin{picture}(50,20)
  \linethickness{0.2mm}
  \put(0,6){\cg \line(1,0){50}}
  \put(0,-4){\cb \line(1,0){10}}
  \put(10,-4){\cg \line(1,0){10}}
  \put(20,-4){\cb \line(1,0){10}} 
  \put(30,-4){\cg \line(1,0){10}}
  \put(40,-4){\cb \line(1,0){10}}
  \put(0,0){{\cre \qbezier(0,1)(30,0)(30,-4)}} 
  \put(0,0){{\cre \qbezier(50,1)(20,0)(20,-4)}} 
  \put(0,0){{\cre \qbezier(10,-4)(25,-15)(40,-4)}}
  \put(10.,-4){\circle{2}}
  \put(20.,-4){\circle*{2}}
  \put(30.,-4){\circle{2}}
  \put(40.,-4){\circle*{2}}
\end{picture} ~~=~~
\begin{picture}(50,10)
  \linethickness{0.2mm}
  \put(0,6){\cg \line(1,0){50}}
  \put(0,-4){\cre \line(1,0){10}}
  \put(10,-4){\cg \line(1,0){10}}
  \put(20,-4){\cre \line(1,0){10}} 
  \put(30,-4){\cg \line(1,0){10}}
  \put(40,-4){\cre \line(1,0){10}}
  \put(0,0){{\cb \qbezier(0,1)(30,0)(30,-4)}} 
  \put(0,0){{\cb \qbezier(50,1)(20,0)(20,-4)}} 
  \put(0,0){{\cb \qbezier(10,-4)(25,-15)(40,-4)}}
  \put(10.,-4){\circle{2}}
  \put(20.,-4){\circle*{2}}
  \put(30.,-4){\circle{2}}
  \put(40.,-4){\circle*{2}}
\end{picture}~~,  ~~\\ 
 &(3W)_{\cb b} = 
 \setlength{\unitlength}{0.1cm}
\begin{picture}(50,20)
  \linethickness{0.2mm}
  \put(0,6){\cb \line(1,0){50}}
  \put(0,-4){\cre \line(1,0){10}}
  \put(10,-4){\cb \line(1,0){10}}
  \put(20,-4){\cre \line(1,0){10}} 
  \put(30,-4){\cb \line(1,0){10}}
  \put(40,-4){\cre \line(1,0){10}}
  \put(0,0){{\cg \qbezier(0,1)(30,0)(30,-4)}} 
  \put(0,0){{\cg \qbezier(50,1)(20,0)(20,-4)}} 
  \put(0,0){{\cg \qbezier(10,-4)(25,-15)(40,-4)}}
  \put(10.,-4){\circle{2}}
  \put(20.,-4){\circle*{2}}
  \put(30.,-4){\circle{2}}
  \put(40.,-4){\circle*{2}}
\end{picture} ~~=~~
\begin{picture}(50,10)
  \linethickness{0.2mm}
  \put(0,6){\cb \line(1,0){50}}
  \put(0,-4){\cg \line(1,0){10}}
  \put(10,-4){\cb \line(1,0){10}}
  \put(20,-4){\cg \line(1,0){10}} 
  \put(30,-4){\cb \line(1,0){10}}
  \put(40,-4){\cg \line(1,0){10}}
  \put(0,0){{\cre \qbezier(0,1)(30,0)(30,-4)}} 
  \put(0,0){{\cre \qbezier(50,1)(20,0)(20,-4)}} 
  \put(0,0){{\cre \qbezier(10,-4)(25,-15)(40,-4)}}
  \put(10.,-4){\circle{2}}
  \put(20.,-4){\circle*{2}}
  \put(30.,-4){\circle{2}}
  \put(40.,-4){\circle*{2}}
\end{picture}~~. ~~
\end{align}\\\\
The subscripts ${\cre r(ed)}$, ${\cg g(reen)}$ and ${\cb b(lue)}$ 
 denote the color which acts trivially.  
Eq. \eqref{K<3} is the linear combination of operators
 whose levels are greater than or equal to 2.  
The levels of the operators in $\tr F^n$ are not always equal to $n$ in such case.  
To be more specific, 
 operators of level $k$ must be included in $\tr F^n$ for some $n \leq k$.

Then, one can observe\footnote{$K_{\rm XXVIII}$ is equivalent 
 to $K_{\rm XXVI}$ except for the replacement of the coloring.}
\begin{align}
 &K_{\rm XXV} = \Tr (3W)_{\cre r}^2 ({\cre 2}) ~~~ \in \tr F^3, 
 \label{XXV}\\
 \checkmark~~~~&K_{\rm XXVI} = \Tr (3W)_{\cre r} ({\cg 2}) ({\cre 2}) ({\cb 2})
 ~~~ \in \tr F^4.
 \label{XXVI} 
 %\checkmark~~~~&K_{\rm XXVIII} = \Tr (3W)_b ({\cre 2}) ({\cb 2}) ({\cg 2}).
\end{align} 
In addition, $K_{\rm XIV}$ is  the secondary operator, 
\begin{align}\label{XIV}
 \checkmark~~~~&K_{\rm XIV} 
 = \Tr ({\cre 2}) ({\cb 2}) ({\cg 2}) ({\cre 2}) ({\cg 2}) ~~~ \in \tr F^5, 
\end{align}
which we can generate in the generalized cut operation already with \eqref{K<2}. 

We now arrive at a conjecture: 
 in order to predict all connected operators at the higher levels,
 all we need to do is to add the new secondary operator at each lower level 
 to $K$ successively.
Then all connected operators at a given level $k$ are included 
 in $\tr F^n (^{\exists}n \leq k)$. 
\begin{equation}
 \text{secondary} = \{K_{3W}, K_{4C}, K_{{\cre 2}{\cre 2}W}, \cdots \}
\end{equation}

\vspace{-5mm}

%%%%%%%%%%%%%%%%%%%%%%%%%%%%%%%%%%%%%%%%%%%%%%%%%%%%%%%%%%%%%%%%%%%%%%%%%%%%%%%%%%
%%%%%%%%%%%%%%%%%%%%%%%%%%%%%%%%%%%%%%%%%%%%%%%%%%%%%%%%%%%%%%%%%%%%%%%%%%%%%%%%%%
\section{Examination at level 6}
%%%%%%%%%%%%%%%%%%%%%%%%%%%%%%%%%%%%%%%%%%%%%%%%%%%%%%%%%%%%%%%%%%%%%%%%%%%%%%%%%%
%%%%%%%%%%%%%%%%%%%%%%%%%%%%%%%%%%%%%%%%%%%%%%%%%%%%%%%%%%%%%%%%%%%%%%%%%%%%%%%%%% 

%%%%%%%%%%%%%%%%%%%%%%%%%%%%%%%%%%%%%%%%%%%%%%%%%%%%%%%%%%%%%%%%%%%%%%%%%%%%%%%%%%
%\subsection{}
%%%%%%%%%%%%%%%%%%%%%%%%%%%%%%%%%%%%%%%%%%%%%%%%%%%%%%%%%%%%%%%%%%%%%%%%%%%%%%%%%%
At level 5, $K_{4C}$ and $K_{{\cre 22}W}$ (also $K_{{\cg 22}W}$ and $K_{{\cb 22}W}$, 
 of course) appear as the new secondary operators. 
Hence, we choose
\begin{equation}
 K_{\leq 4} = K_{\cre{2}} + K_{\cg{2}} + K_{\cb{2}} + K_{3W} + 
 K_{4C} + K_{{\cre 22}W} + K_{{\cg 22}W} + K_{{\cb 22}W}. 
\end{equation}
We then have 
\begin{equation}
 \p\bar{\p} K_{4C} = 4 \biggl\{ (4C) + (4C)_{\cre r} + (4C)_{\cg g} + (4C)_{\cb b} \biggl\},
\end{equation}
where \vspace{-10mm}
\begin{gather}
 (4C)_{\cre r} = 
 \setlength{\unitlength}{0.1cm}
\begin{picture}(30,20)
  \linethickness{0.2mm}
  \put(0,12){\cre \line(1,0){30}}
  \put(0,-8){\cb \line(1,0){10}}
  \put(10,-8){\cre \line(1,0){10}}
  \put(20,-8){\cb \line(1,0){10}} 
  \put(10,0){\cre \line(1,0){10}}
  \put(10,8){\cre \line(1,0){10}}
  \put(10,-8){\cg \line(0,1){8}}
  \put(20,-8){\cg \line(0,1){8}}
  \put(10,0){\cb \line(0,1){8}}
  \put(20,0){\cb \line(0,1){8}}
  \put(10,8){\cg \line(-1,0){10}}
  \put(20,8){\cg \line(1,0){10}}  
  \put(10,-8){\circle{2}}
  \put(20,-8){\circle*{2}}
  \put(10,0){\circle*{2}}
  \put(20,0){\circle{2}}
  \put(10,8){\circle{2}}
  \put(20,8){\circle*{2}}
\end{picture} ~~,~~~
(4C)_{\cg g} = 
\begin{picture}(30,20)
  \linethickness{0.2mm}
  \put(0,12){\cg \line(1,0){30}}
  \put(0,-8){\cre \line(1,0){10}}
  \put(10,-8){\cg \line(1,0){10}}
  \put(20,-8){\cre \line(1,0){10}} 
  \put(10,0){\cg \line(1,0){10}}
  \put(10,8){\cg \line(1,0){10}}
  \put(10,-8){\cb \line(0,1){8}}
  \put(20,-8){\cb \line(0,1){8}}
  \put(10,0){\cre \line(0,1){8}}
  \put(20,0){\cre \line(0,1){8}}
  \put(10,8){\cb \line(-1,0){10}}
  \put(20,8){\cb \line(1,0){10}}  
  \put(10,-8){\circle{2}}
  \put(20,-8){\circle*{2}}
  \put(10,0){\circle*{2}}
  \put(20,0){\circle{2}}
  \put(10,8){\circle{2}}
  \put(20,8){\circle*{2}}
\end{picture} ~~,~~~
(4C)_{\cb b} = 
\begin{picture}(30,20)
  \linethickness{0.2mm}
  \put(0,12){\cb \line(1,0){30}}
  \put(0,-8){\cg \line(1,0){10}}
  \put(10,-8){\cb \line(1,0){10}}
  \put(20,-8){\cg \line(1,0){10}} 
  \put(10,0){\cb \line(1,0){10}}
  \put(10,8){\cb \line(1,0){10}}
  \put(10,-8){\cre \line(0,1){8}}
  \put(20,-8){\cre \line(0,1){8}}
  \put(10,0){\cg \line(0,1){8}}
  \put(20,0){\cg \line(0,1){8}}
  \put(10,8){\cre \line(-1,0){10}}
  \put(20,8){\cre \line(1,0){10}}  
  \put(10,-8){\circle{2}}
  \put(20,-8){\circle*{2}}
  \put(10,0){\circle*{2}}
  \put(20,0){\circle{2}}
  \put(10,8){\circle{2}}
  \put(20,8){\circle*{2}}
\end{picture}
\end{gather} \vspace{-20mm}
\begin{gather} 
 (4C) = 
 \setlength{\unitlength}{0.1cm}
\begin{picture}(40,30)
  \linethickness{0.2mm}
  \put(0,-6){\cre \line(1,0){10}}
  \put(10,-6){\cg \line(1,0){10}}
  \put(20,-6){\cre \line(1,0){10}} 
  \put(10,-6){\cb \line(0,1){12}}
  \put(30,-6){\cb \line(0,1){12}}
  \put(10,6){\cre \line(1,0){10}}
  \put(20,6){\cg \line(1,0){10}}
  \put(30,6){\cre \line(1,0){10}} 
  \put(0,0){{\cb \qbezier(20,6)(20,2)(0,2)}} 
  \put(0,0){{\cb \qbezier(40,-2)(20,-2)(20,-6)}}
  \put(0,0){{\cg \qbezier(10,6)(10,10)(40,10)}}
  \put(0,0){{\cg \qbezier(30,-6)(30,-10)(0,-10)}}
  \put(10.,-6){\circle{2}}
  \put(20.,-6){\circle*{2}}
  \put(30.,-6){\circle{2}}
  \put(10.,6){\circle*{2}}
  \put(20.,6){\circle{2}}
  \put(30.,6){\circle*{2}}
\end{picture} ~~, ~~\\\notag
\end{gather}
and 
\begin{equation}
 \p\bar{\p} K_{{\cre 2}{\cre 2}W} 
 = 4 \biggl\{ ({\cre 22}W) + ({\cre 22}W)_{\cre r} + ({\cre 22}W)_{\cg g} 
 + ({\cre 22}W)_{\cb b} \biggl\},
\end{equation}
where \vspace{-10mm}
\begin{gather}
 ({\cre 22}W)_{\cre r} = 
\setlength{\unitlength}{0.1cm}
\begin{picture}(30,25)
  \linethickness{0.2mm}
  \put(0,12){\cre \line(1,0){30}}
  \put(0,-8){\cb \line(1,0){10}}
  \put(10,-8){\cre \line(1,0){10}}
  \put(20,-8){\cg \line(1,0){10}} 
  \put(10,0){\cre \line(1,0){10}}
  \put(10,8){\cre \line(1,0){10}}
  \put(10,-8){\cg \line(0,1){8}}
  \put(20,-8){\cb \line(0,1){8}}
  \put(10,0){\cb \line(0,1){8}}
  \put(20,0){\cg \line(0,1){8}}
  \put(10,8){\cg \line(-1,0){10}}
  \put(20,8){\cb \line(1,0){10}}  
  \put(10,-8){\circle{2}}
  \put(20,-8){\circle*{2}}
  \put(10,0){\circle*{2}}
  \put(20,0){\circle{2}}
  \put(10,8){\circle{2}}
  \put(20,8){\circle*{2}}
\end{picture} ~~,~~~ \\
({\cre 22}W)_{\cg g} = 
 \setlength{\unitlength}{0.1cm}
\begin{picture}(30,25)
  \linethickness{0.2mm}
  \put(0,12){\cg \line(1,0){30}}
  \put(0,-8){\cre \line(1,0){10}}
  \put(10,-8){\cb \line(1,0){10}}
  \put(20,-8){\cre \line(1,0){10}} 
  \put(10,0){\cb \line(1,0){10}}
  \put(10,8){\cg \line(1,0){10}}
  \put(10,-8){\cg \line(0,1){8}}
  \put(20,-8){\cg \line(0,1){8}}
  \put(10,0){\cre \line(0,1){8}}
  \put(20,0){\cre \line(0,1){8}}
  \put(10,8){\cb \line(-1,0){10}}
  \put(20,8){\cb \line(1,0){10}}  
  \put(10,-8){\circle{2}}
  \put(20,-8){\circle*{2}}
  \put(10,0){\circle*{2}}
  \put(20,0){\circle{2}}
  \put(10,8){\circle{2}}
  \put(20,8){\circle*{2}}
\end{picture} ~~, ~~~~~~~
({\cre 22}W)_{\cb b} = 
 \setlength{\unitlength}{0.1cm}
\begin{picture}(30,25)
  \linethickness{0.2mm}
  \put(0,12){\cb \line(1,0){30}}
  \put(0,-8){\cre \line(1,0){10}}
  \put(10,-8){\cg \line(1,0){10}}
  \put(20,-8){\cre \line(1,0){10}} 
  \put(10,0){\cg \line(1,0){10}}
  \put(10,8){\cb \line(1,0){10}}
  \put(10,-8){\cb \line(0,1){8}}
  \put(20,-8){\cb \line(0,1){8}}
  \put(10,0){\cre \line(0,1){8}}
  \put(20,0){\cre \line(0,1){8}}
  \put(10,8){\cg \line(-1,0){10}}
  \put(20,8){\cg \line(1,0){10}}  
  \put(10,-8){\circle{2}}
  \put(20,-8){\circle*{2}}
  \put(10,0){\circle*{2}}
  \put(20,0){\circle{2}}
  \put(10,8){\circle{2}}
  \put(20,8){\circle*{2}}
\end{picture} ~~, \\
 ({\cre 22}W) = 
 \setlength{\unitlength}{0.1cm}
\begin{picture}(40,30)
  \linethickness{0.2mm}
  \put(0,-6){\cre \line(1,0){10}}
  \put(10,-6){\cg \line(1,0){10}}
  \put(20,-6){\cre \line(1,0){10}} 
  \put(10,-6){\cb \line(0,1){12}}
  \put(30,-6){\cg \line(0,1){12}}
  \put(10,6){\cre \line(1,0){10}}
  \put(20,6){\cb \line(1,0){10}}
  \put(30,6){\cre \line(1,0){10}} 
  \put(0,0){{\cg \qbezier(20,6)(20,2)(0,2)}} 
  \put(0,0){{\cb \qbezier(40,-2)(20,-2)(20,-6)}}
  \put(0,0){{\cg \qbezier(10,6)(10,10)(40,10)}}
  \put(0,0){{\cb \qbezier(30,-6)(30,-10)(0,-10)}}
  \put(10.,-6){\circle{2}}
  \put(20.,-6){\circle*{2}}
  \put(30.,-6){\circle{2}}
  \put(10.,6){\circle*{2}}
  \put(20.,6){\circle{2}}
  \put(30.,6){\circle*{2}}
\end{picture} ~~. ~~
\end{gather}\\\\
We checked by direct inspection that all operators at level 6 are included in 
 $\tr F^n (^\exists n \leq 6)$ with $K_{\leq 4}$. 
 We plan to elaborate upon this in the future.   
In particular, we found 10 independent secondary operators at level 6 
 up to the coloring,   \\ 
 
\vspace*{-30mm}
\begin{align}
 \Tr ({\cre 2}) ({\cg 2}) ({\cre 2}) ({\cg 2}) ({\cre 2}) ({\cg 2})
 &=~~~~~
 \begin{picture}(90,80)
  \linethickness{0.2mm}
  \put(-4,0){
  {\cre{\qbezier(20,34)(10,17)(0,0)}
  {\qbezier(20,-34)(40,-34)(60,-34)}
  {\qbezier(80,0)(70,17)(60,34)}
 {\qbezier(20,0)(25,8.5)(30,17)}
 {\qbezier(50,17)(55,8.5)(60,0)}
 {\qbezier(30,-17)(40,-17)(50,-17)}
 }}
  \put(-4,0){
  {\cg 
  {\qbezier(0,0)(10,-17)(20,-34)}
  {\qbezier(60,-34)(70,-17)(80,0)}
  {\qbezier(60,34)(40,34)(20,34)}
  {\qbezier(20,0)(25,-8.5)(30,-17)}
  {\qbezier(50,-17)(55,-8.5)(60,0)}
  {\qbezier(30,17)(40,17)(50,17)}
 }}
  \put(-4,0){
  {\cb 
  {\qbezier(20,34)(25,25.5)(30,17)}
  {\qbezier(0,0)(10,0)(20,0)}
  {\qbezier(50,17)(55,25.5)(60,34)}
  {\qbezier(60,0)(70,0)(80,0)}
  {\qbezier(50,-17)(55,-25.5)(60,-34)} 
  {\qbezier(20,-34)(25,-25.5)(30,-17)} 
  }}
  \put(20,34){\circle{6}}
  \put(20,-34){\circle{6}}
  \put(80,0){\circle{6}}
  \put(0,0){\circle*{6}}
  \put(60,-34){\circle*{6}}
  \put(60,34){\circle*{6}}
  \put(30,17){\circle*{6}}
  \put(50,17){\circle{6}}
  \put(60,0){\circle*{6}}
  \put(50,-17){\circle{6}}
  \put(30,-17){\circle*{6}}
  \put(20,0){\circle{6}}
  \end{picture} , 
 \label{1:level6}\\
 \Tr ({\cre 2}) ({\cg 2}) ({\cre 2}) ({\cg 2}) ({\cre 2}) ({\cb 2})
  &=~~~~~
 \begin{picture}(90,80)
  \linethickness{0.2mm}
  \put(-4,0){
  {\cre
  {\qbezier(0,10)(0,0)(0,-10)}
  {\qbezier(16,-24)(24,-31)(32,-38)}
  {\qbezier(52,-38)(60,-31)(68,-24)}
  {\qbezier(84,-10)(84,0)(84,10)}
  {\qbezier(68,24)(60,31)(52,38)}
  {\qbezier(32,38)(24,31)(16,24)}
 }}
  \put(-4,0){
  {\cg 
  {\qbezier(0,10)(8,17)(16,24)}
  {\qbezier(32,38)(42,38)(52,38)}
  {\qbezier(68,24)(76,17)(84,10)}
  {\qbezier(0,-10)(8,-17)(16,-24)}
  {\qbezier(32,-38)(42,-38)(52,-38)}
  {\qbezier(68,-24)(76,-17)(84,-10)}
 }}
  \put(-4,0){
  {\cb 
  {\qbezier(0,10)(42,10)(84,10)}
  {\qbezier(16,24)(42,24)(68,24)}
  {\qbezier(0,-10)(42,-10)(84,-10)}
  {\qbezier(16,-24)(42,-24)(68,-24)}
  {\qbezier(32,38)(32,0)(32,-38)}
  {\qbezier(52,38)(52,0)(52,-38)}
 }}
  \put(0,10){\circle{6}}
  \put(16,24){\circle*{6}}
  \put(32,38){\circle{6}}
  \put(52,38){\circle*{6}}
  \put(68,24){\circle{6}}
  \put(84,10){\circle*{6}}
  \put(0,-10){\circle*{6}}
  \put(16,-24){\circle{6}}
  \put(32,-38){\circle*{6}}
  \put(52,-38){\circle{6}}
  \put(68,-24){\circle*{6}}
  \put(84,-10){\circle{6}}
  \end{picture} ,  
 \label{2:level6}\\
 \Tr ({\cre 2}) ({\cg 2}) ({\cre 2}) ({\cb 2}) ({\cg 2}) ({\cb 2})
  &=~~~~~~
 \begin{picture}(90,80)
  \linethickness{0.2mm}
  \put(-4,0){
  {\cre
  {\qbezier(0,14)(12,24)(24,34)}
  {\qbezier(52,34)(64,24)(76,14)}
  {\qbezier(24,10)(38,10)(52,10)}
  {\qbezier(0,-14)(12,-24)(24,-34)}
  {\qbezier(52,-34)(64,-24)(76,-14)}
  {\qbezier(24,-10)(38,-10)(52,-10)}
 }}
  \put(-4,0){
  {\cg 
  {\qbezier(24,34)(38,34)(52,34)}
  {\qbezier(24,-34)(38,-34)(52,-34)}
  {\qbezier(0,14)(0,0)(0,-14)}
  {\qbezier(76,14)(76,0)(76,-14)}
  {\qbezier(24,10)(24,0)(24,-10)}
  {\qbezier(52,10)(52,0)(52,-10)}
 }}
  \put(-4,0){
  {\cb 
  {\qbezier(24,34)(24,22)(24,10)}
  {\qbezier(52,34)(52,22)(52,10)}
  {\qbezier(24,-34)(24,-22)(24,-10)}
  {\qbezier(52,-34)(52,-22)(52,-10)}
  {\qbezier(0,14)(38,30)(76,14)}
  {\qbezier(0,-14)(38,-30)(76,-14)}
  }}
  \put(0,14){\circle*{6}}
  \put(24,34){\circle{6}}
  \put(52,34){\circle*{6}}
  \put(76,14){\circle{6}}
  \put(0,-14){\circle{6}}
  \put(24,-34){\circle*{6}}
  \put(52,-34){\circle{6}}
  \put(76,-14){\circle*{6}}
  \put(24,10){\circle*{6}}
  \put(52,10){\circle{6}}
  \put(24,-10){\circle{6}}
  \put(52,-10){\circle*{6}}
  \end{picture} ,  
 \label{3:level6} \\
 \Tr ({\cre 2}) ({\cg 2}) ({\cb 2}) ({\cre 2}) ({\cg 2}) ({\cb 2})
  &=~~~~~
 \begin{picture}(90,80)
  \linethickness{0.2mm}
  \put(-4,0){
  {\cre{\qbezier(20,34)(10,17)(0,0)}
  {\qbezier(20,-34)(40,-34)(60,-34)}
  {\qbezier(80,0)(70,17)(60,34)}
 {\qbezier(20,0)(25,8.5)(30,17)}
 {\qbezier(50,17)(55,8.5)(60,0)}
 {\qbezier(30,-17)(40,-17)(50,-17)}
 }}
  \put(-4,0){
  {\cg 
  {\qbezier(0,0)(10,-17)(20,-34)}
  {\qbezier(60,-34)(70,-17)(80,0)}
  {\qbezier(60,34)(40,34)(20,34)}
  {\qbezier(20,0)(25,-8.5)(30,-17)}
  {\qbezier(50,-17)(55,-8.5)(60,0)}
  {\qbezier(30,17)(40,17)(50,17)}
 }}
  \put(-4,0){
  {\cb 
  {\qbezier(20,34)(25,25.5)(30,17)}
  {\qbezier(0,0)(40,60)(80,0)}
  {\qbezier(50,17)(55,25.5)(60,34)}
  {\qbezier(60,0)(40,0)(20,0)}
  {\qbezier(50,-17)(55,-25.5)(60,-34)} 
  {\qbezier(20,-34)(25,-25.5)(30,-17)} 
  }}
  \put(20,34){\circle{6}}
  \put(20,-34){\circle{6}}
  \put(80,0){\circle{6}}
  \put(0,0){\circle*{6}}
  \put(60,-34){\circle*{6}}
  \put(60,34){\circle*{6}}
  \put(30,17){\circle*{6}}
  \put(50,17){\circle{6}}
  \put(60,0){\circle*{6}}
  \put(50,-17){\circle{6}}
  \put(30,-17){\circle*{6}}
  \put(20,0){\circle{6}}
  \end{picture} ,  
 \label{4:level6}\\
 \Tr (3W)_{\cre r} ({\cg 2}) ({\cb 2}) ({\cg 2}) ({\cb 2})
  &=~~~~~~
 \begin{picture}(90,80)
  \linethickness{0.2mm}
  \put(-4,0){
  {\cre
  {\qbezier(0,14)(12,24)(24,34)}
  {\qbezier(52,34)(64,24)(76,14)}
  {\qbezier(0,-14)(12,-24)(24,-34)}
  {\qbezier(52,-34)(64,-24)(76,-14)}
  {\qbezier(30,0)(38,7)(46,14)}
  {\qbezier(46,-14)(54,-7)(62,0)}
 }}
  \put(-4,0){
  {\cg 
  {\qbezier(24,34)(38,34)(52,34)}
  {\qbezier(24,-34)(38,-34)(52,-34)}
  {\qbezier(0,14)(0,0)(0,-14)}
  {\qbezier(76,14)(76,0)(76,-14)}
  {\qbezier(30,0)(46,0)(62,0)}
  {\qbezier(46,14)(46,0)(46,-14)}
 }}
  \put(-4,0){
  {\cb 
  {\qbezier(0,14)(38,36)(76,14)}
  {\qbezier(24,34)(8,0)(24,-34)}
  {\qbezier(52,34)(49,24)(46,14)}
  {\qbezier(0,-14)(15,-7)(30,0)}
  {\qbezier(52,-34)(49,-24)(46,-14)}
  {\qbezier(76,-14)(69,-7)(62,0)}
  }}
  \put(0,14){\circle*{6}}
  \put(24,34){\circle{6}}
  \put(52,34){\circle*{6}}
  \put(76,14){\circle{6}}
  \put(0,-14){\circle{6}}
  \put(24,-34){\circle*{6}}
  \put(52,-34){\circle{6}}
  \put(76,-14){\circle*{6}}
  \put(46,14){\circle{6}}
  \put(30,0){\circle*{6}}
  \put(46,-14){\circle*{6}}
  \put(62,0){\circle{6}}
  \end{picture} ,
 \label{5:level6}\\
  \Tr (3W)_{\cre r} ({\cg 2}) ({\cre 2}) ({\cg 2}) ({\cb 2})
  &=~~~~~
 \begin{picture}(90,80)
  \linethickness{0.2mm}
  \put(-4,0){
  {\cre
  {\qbezier(0,10)(0,0)(0,-10)}
  {\qbezier(16,-24)(24,-31)(32,-38)}
  {\qbezier(52,-38)(60,-31)(68,-24)}
  {\qbezier(84,-10)(84,0)(84,10)}
  {\qbezier(68,24)(60,31)(52,38)}
  {\qbezier(32,38)(24,31)(16,24)}
 }}
  \put(-4,0){
  {\cg 
  {\qbezier(0,10)(8,17)(16,24)}
  {\qbezier(32,38)(42,38)(52,38)}
  {\qbezier(68,24)(76,17)(84,10)}
  {\qbezier(0,-10)(8,-17)(16,-24)}
  {\qbezier(32,-38)(42,-38)(52,-38)}
  {\qbezier(68,-24)(76,-17)(84,-10)}
 }}
  \put(-4,0){
  {\cb 
  {\qbezier(0,10)(34,-7)(68,-24)}
  {\qbezier(16,24)(42,24)(68,24)}
  {\qbezier(32,38)(26,4)(0,-10)}
  {\qbezier(52,38)(52,0)(52,-38)}
  {\qbezier(84,10)(42,24)(16,-24)}
  {\qbezier(32,-38)(35,-5)(84,-10)}
 }}
  \put(0,10){\circle{6}}
  \put(16,24){\circle*{6}}
  \put(32,38){\circle{6}}
  \put(52,38){\circle*{6}}
  \put(68,24){\circle{6}}
  \put(84,10){\circle*{6}}
  \put(0,-10){\circle*{6}}
  \put(16,-24){\circle{6}}
  \put(32,-38){\circle*{6}}
  \put(52,-38){\circle{6}}
  \put(68,-24){\circle*{6}}
  \put(84,-10){\circle{6}}
  \end{picture} ,   
 \label{6:level6}\\
 \Tr (3W)_{\cre r} ({\cg 2}) (3W)_{\cre r} ({\cb 2})
  &=~~~~~
 \begin{picture}(90,80)
  \linethickness{0.2mm}
  \put(-4,0){
  {\cre
  {\qbezier(0,10)(0,0)(0,-10)}
  {\qbezier(16,-24)(24,-31)(32,-38)}
  {\qbezier(52,-38)(60,-31)(68,-24)}
  {\qbezier(84,-10)(84,0)(84,10)}
  {\qbezier(68,24)(60,31)(52,38)}
  {\qbezier(32,38)(24,31)(16,24)}
 }}
  \put(-4,0){
  {\cg 
  {\qbezier(0,10)(8,17)(16,24)}
  {\qbezier(32,38)(42,38)(52,38)}
  {\qbezier(68,24)(76,17)(84,10)}
  {\qbezier(0,-10)(8,-17)(16,-24)}
  {\qbezier(32,-38)(42,-38)(52,-38)}
  {\qbezier(68,-24)(76,-17)(84,-10)}
 }}
  \put(-4,0){
  {\cb 
  {\qbezier(0,10)(28,2)(32,-38)}
  {\qbezier(16,24)(42,9)(68,24)}
  {\qbezier(32,38)(28,2)(0,-10)}
  {\qbezier(52,38)(56,2)(84,-10)}
  {\qbezier(84,10)(56,2)(52,-38)}
  {\qbezier(16,-24)(42,-9)(68,-24)}
 }}
  \put(0,10){\circle{6}}
  \put(16,24){\circle*{6}}
  \put(32,38){\circle{6}}
  \put(52,38){\circle*{6}}
  \put(68,24){\circle{6}}
  \put(84,10){\circle*{6}}%
  \put(0,-10){\circle*{6}}
  \put(16,-24){\circle{6}}
  \put(32,-38){\circle*{6}}
  \put(52,-38){\circle{6}}
  \put(68,-24){\circle*{6}}
  \put(84,-10){\circle{6}}
  \end{picture} ,  \label{7:level6} %\\
 \end{align} \vspace*{-20mm}
 \begin{align}
 \Tr (4C) ({\cg 2}) ({\cre 2}) ({\cg 2})
  &=~~~~~
 \begin{picture}(90,80)
  \linethickness{0.2mm}
  \put(-4,0){
  {\cre
  {\qbezier(0,10)(0,0)(0,-10)}
  {\qbezier(16,-24)(24,-31)(32,-38)}
  {\qbezier(52,-38)(60,-31)(68,-24)}
  {\qbezier(84,-10)(84,0)(84,10)}
  {\qbezier(68,24)(60,31)(52,38)}
  {\qbezier(32,38)(24,31)(16,24)}
 }}
  \put(-4,0){
  {\cg 
  {\qbezier(0,10)(8,17)(16,24)}
  {\qbezier(32,38)(42,38)(52,38)}
  {\qbezier(68,24)(76,17)(84,10)}
  {\qbezier(0,-10)(8,-17)(16,-24)}
  {\qbezier(32,-38)(42,-38)(52,-38)}
  {\qbezier(68,-24)(76,-17)(84,-10)}
 }}
  \put(-4,0){
  {\cb 
  {\qbezier(0,10)(42,10)(84,10)}
  {\qbezier(16,24)(34,-7)(52,-38)}
  {\qbezier(32,38)(50,7)(68,-24)}
  {\qbezier(52,38)(34,7)(16,-24)}
  {\qbezier(68,24)(50,2)(32,-38)}
  {\qbezier(0,-10)(42,-10)(84,-10)}
 }}
  \put(0,10){\circle{6}}
  \put(16,24){\circle*{6}}
  \put(32,38){\circle{6}}
  \put(52,38){\circle*{6}}
  \put(68,24){\circle{6}}
  \put(84,10){\circle*{6}}%
  \put(0,-10){\circle*{6}}
  \put(16,-24){\circle{6}}
  \put(32,-38){\circle*{6}}
  \put(52,-38){\circle{6}}
  \put(68,-24){\circle*{6}}
  \put(84,-10){\circle{6}}
  \end{picture} ,   
 \label{8:level6}\\
 \Tr (4C) ({\cre 22}W)
  &=~~~~~
 \begin{picture}(90,80)
  \linethickness{0.2mm}
  \put(-4,0){
  {\cre
  {\qbezier(0,10)(0,0)(0,-10)}
  {\qbezier(16,-24)(24,-31)(32,-38)}
  {\qbezier(52,-38)(60,-31)(68,-24)}
  {\qbezier(84,-10)(84,0)(84,10)}
  {\qbezier(68,24)(60,31)(52,38)}
  {\qbezier(32,38)(24,31)(16,24)}
 }}
  \put(-4,0){
  {\cg 
  {\qbezier(0,10)(8,17)(16,24)}
  {\qbezier(32,38)(42,38)(52,38)}
  {\qbezier(68,24)(76,17)(84,10)}
  {\qbezier(0,-10)(8,-17)(16,-24)}
  {\qbezier(32,-38)(42,-38)(52,-38)}
  {\qbezier(68,-24)(76,-17)(84,-10)}
 }}
  \put(-4,0){
  {\cb 
  {\qbezier(0,10)(34,-7)(68,-24)}
  {\qbezier(16,24)(34,-7)(52,-38)}
  {\qbezier(32,38)(51,11)(84,10)}
  {\qbezier(0,-10)(34,7)(68,24)}
  {\qbezier(16,-24)(34,7)(52,38)}
  {\qbezier(32,-38)(51,-11)(84,-10)}
 }}
  \put(0,10){\circle{6}}
  \put(16,24){\circle*{6}}
  \put(32,38){\circle{6}}
  \put(52,38){\circle*{6}}
  \put(68,24){\circle{6}}
  \put(84,10){\circle*{6}}%
  \put(0,-10){\circle*{6}}
  \put(16,-24){\circle{6}}
  \put(32,-38){\circle*{6}}
  \put(52,-38){\circle{6}}
  \put(68,-24){\circle*{6}}
  \put(84,-10){\circle{6}}
  \end{picture} , 
 \label{9:level6}\\
 \Tr ({\cg 22} W) ({\cb 22}W)
  &=~~~~~
 \begin{picture}(90,80)
  \linethickness{0.2mm}
  \put(-4,0){
  {\cre
  {\qbezier(0,10)(0,0)(0,-10)}
  {\qbezier(16,-24)(24,-31)(32,-38)}
  {\qbezier(52,-38)(60,-31)(68,-24)}
  {\qbezier(84,-10)(84,0)(84,10)}
  {\qbezier(68,24)(60,31)(52,38)}
  {\qbezier(32,38)(24,31)(16,24)}
 }}
  \put(-4,0){
  {\cg 
  {\qbezier(0,10)(8,17)(16,24)}
  {\qbezier(32,38)(42,38)(52,38)}
  {\qbezier(68,24)(76,17)(84,10)}
  {\qbezier(0,-10)(8,-17)(16,-24)}
  {\qbezier(32,-38)(42,-38)(52,-38)}
  {\qbezier(68,-24)(76,-17)(84,-10)}
 }}
  \put(-4,0){
  {\cb 
  {\qbezier(0,10)(28,2)(32,-38)}
  {\qbezier(16,24)(42,9)(68,24)}
  {\qbezier(32,38)(50,7)(68,-24)}
  {\qbezier(52,38)(34,7)(16,-24)}
  {\qbezier(84,10)(56,2)(52,-38)}
  {\qbezier(0,-10)(42,-10)(84,-10)}
 }}
  \put(0,10){\circle{6}}
  \put(16,24){\circle*{6}}
  \put(32,38){\circle{6}}
  \put(52,38){\circle*{6}}
  \put(68,24){\circle{6}}
  \put(84,10){\circle*{6}}%
  \put(0,-10){\circle*{6}}
  \put(16,-24){\circle{6}}
  \put(32,-38){\circle*{6}}
  \put(52,-38){\circle{6}}
  \put(68,-24){\circle*{6}}
  \put(84,-10){\circle{6}}
  \end{picture} .  
 \label{10:level6} 
\end{align}\\\\

The secondary operators can be constructed by the appropriate product of    
 the objects $({\cre 2})$, $({\cg 2})$, $({\cb 2})$, $(3W)_{\cre r}$ and so on
 and the trace ``$\Tr$".

%%%%%%%%%%%%%%%%%%%%%%%%%%%%%%%%%%%%%%%%%%%%%%%%%%%%%%%%%%%%%%%%%%%%%%%%%%%%%%%%%%
%%%%%%%%%%%%%%%%%%%%%%%%%%%%%%%%%%%%%%%%%%%%%%%%%%%%%%%%%%%%%%%%%%%%%%%%%%%%%%%%%%
\section{Construction of the secondary operators}
%%%%%%%%%%%%%%%%%%%%%%%%%%%%%%%%%%%%%%%%%%%%%%%%%%%%%%%%%%%%%%%%%%%%%%%%%%%%%%%%%%
%%%%%%%%%%%%%%%%%%%%%%%%%%%%%%%%%%%%%%%%%%%%%%%%%%%%%%%%%%%%%%%%%%%%%%%%%%%%%%%%%% 

In the previous section, we have seen that, up to level 6, 
 all operators appear as the constituents of $\tr F^n (n \leq 6)$ .     
In particular, the secondary operators are constructed by the trace 
 of the product of the ingredients $({\cre 2})$, $({\cg 2})$, $({\cb 2})$, 
 $(3W)_{\cre r}$ and so on. 
Then a natural question arises  
as to  what combinations of these ingredients the secondary operators consist of.
Unfortunately, we do not have an complete answer. 
However, there may be some rules
 as to the correspondence between a ``word" and each of the secondary operators. 

\begin{figure}[h]
 \begin{center}
  \includegraphics[height=3.5cm]{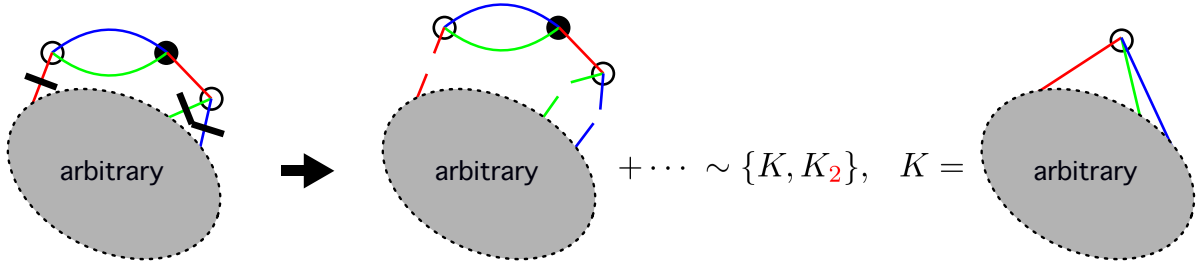}
 \end{center}
\caption{Any level $n$ operator with loop is obtained 
 by the join operation of the level $n-1$ operator $K$ with $K_2$. 
}\label{loop}
\end{figure}

The join operation $\{K, K_2\}$  is the following operation 
 in the pictorial representation:  
 one of the white circles (resp. the black dots) in $K$ (resp. $K_2$) are removed  
 and then the open lines with the same color are connected with each other. 
Thus if an operator can be split into two sub-diagrams 
 by cutting one line per each color, 
 it appears in the join operation pyramid. 
From this fact, the following corollary follows at once: 
Since the operators including the loop 
\setlength{\unitlength}{0.5mm}
\begin{picture}(25,10)
  \linethickness{0.2mm}
  %\put(-5,0){ \line(1,0){5}}
  %\put(21,0){ \line(1,0){5}}
  \put(-4,0){ \line(1,0){21}} 
  \put(-20,4){{ \qbezier(30,2)(20,1)(20,-4)}} 
  \put(0,0){\circle*{3.5}}
  \put(10,4){{ \qbezier(0,2)(10,1)(10,-4)}} 
  \put(20,0){\circle{3.5}}
\end{picture} 
 can always be split into two diagrams as shown in Fig. \ref{loop},  
 they are obtained by the join operation. 
%------------------------FIGURE----------------------------------
%----------------------------------------------------------------

Since the existence of the loops in a diagram means 
 that the operator can be obtained by the join operation, 
the diagram that includes $({\cre 2})'$, $({\cg 2})'$ or $({\cb 2})'$ 
 does not correspond to the secondary operators by construction. 

Moreover, each of the ingredients $({\cre 2})$, $({\cg 2})$ and $({\cb 2})$
 can not be repeated if these have the same color    
because loops are always generated in such cases.     
For example, $({\cre 2}) ({\cre 2})$ generates one green-blue loop,  
\begin{align}
 ({\cre 2}) ({\cre 2}) = 
 \setlength{\unitlength}{0.1cm}
\begin{picture}(60,10)
  \linethickness{0.2mm}
  \put(0,6){\cre \line(1,0){30}}
  \put(0,-4){\cg \line(1,0){10}}
  \put(10,-4){\cre \line(1,0){10}}
  \put(20,-4){\cg \line(1,0){10}} 
  \put(0,0){{\cb \qbezier(0,1)(10,0)(10,-4)}} 
  \put(0,0){{\cb \qbezier(30,1)(20,0)(20,-4)}} 
  \put(10.,-4){\circle{2}}
  \put(20.,-4){\circle*{2}}
\put(30,6){\cre \line(1,0){30}}
  \put(30,-4){\cg \line(1,0){10}}
  \put(40,-4){\cre \line(1,0){10}}
  \put(50,-4){\cg \line(1,0){10}} 
  \put(30,0){{\cb \qbezier(0,1)(10,0)(10,-4)}} 
  \put(30,0){{\cb \qbezier(30,1)(20,0)(20,-4)}} 
  \put(40.,-4){\circle{2}}
  \put(50.,-4){\circle*{2}}
\end{picture}~~ .
\end{align}\\
This restriction on the repeated use of the ingredients with the same coloring
 is extended to the objects with subscript ${\cre r}$, ${\cg g}$, ${\cb b}$, 
  such as $(3W)_{\cre r}$. 
For example, $(3W)_{\cre  r}({\cre 2})$ can always be split
 by cutting the lines depicted by the thick black lines as follows:    
\begin{align}
 (3W)_{\cre r} ({\cre 2}) =  
 \setlength{\unitlength}{0.1cm}
\begin{picture}(70,10)
  \linethickness{0.2mm}
  \put(0,6){\cre \line(1,0){70}}
  \put(0,0){{\cb \qbezier(0,1)(30,0)(30,-4)}} 
  \put(0,0){{\cb \qbezier(20,-4)(35,6)(50,-4)}} 
  \put(0,0){{\cb \qbezier(10,-4)(25,-14)(40,-4)}} 
  \put(0,0){{\cb \qbezier(60,-4)(60,0)(70,1)}} 
  \put(0,-4){\cg \line(1,0){10}}
  \put(10,-4){\cre \line(1,0){10}}
  \put(20,-4){\cg \line(1,0){10}}
  \put(30,-4){\cre \line(1,0){10}}
  \put(40,-4){\cg \line(1,0){10}}
  \put(50,-4){\cre \line(1,0){10}}
  \put(60,-4){\cg \line(1,0){10}}
  \put(10.,-4){\circle{2}}
  \put(20.,-4){\circle*{2}}
  \put(30.,-4){\circle{2}}
  \put(40.,-4){\circle*{2}}
  \put(50.,-4){\circle{2}}
  \put(60.,-4){\circle*{2}}
  \linethickness{3pt}
  \put(5,-7){\line(0,1){10}}
  \linethickness{3pt}
  \put(55,-7){\line(0,1){5.5}}
\end{picture}~~.  
\end{align}\\

In fact, \eqref{3W}, \eqref{4C}, \eqref{22W}, \eqref{XXVI}, 
 \eqref{XXV}, \eqref{XIV},  \eqref{1:level6}-\eqref{7:level6} 
 satisfy these restrictions.  
However, 
we have not been able to formulate rules for \eqref{8:level6}-\eqref{10:level6} 
 by the computation up to level 6.  
In addition, we have seen to cases in which 
 different ``words" yield the same operator.
Despite these incompleteness of the currently constructed rules, 
 in principle, our procedure successfully generates all secondary operators
 level by level recursively.

%%%%%%%%%%%%%%%%  acknowledgements %%%%%%%%%%%%%%%%%%%%%%
\section*{Acknowledgements}
%%%%%%%%%%%%%%%%%%%%%%%%%%%%%%%%%%%%%%%%%%%%%%%%
One of the authors (H. I.) acknowledges Andrei Mironov and Alyosha Morozov for collaboration  of \cite{IMM1703,IMM1704,IMM1710,IMM1808}.

%%%%%%%%%%%%%%%%%%%%%%%%%%%%%%%%%%%%%%%%%%%%%%%%
%\bibliographystyle{arxiv}
%\bibliography{tensor}
%%%%%%%%%%%%%%%%%%%%%%%%%%%%%%%%%%%%%%%%%%%%%%%%

\end{document}